\documentclass[a4paper,11pt]{article}
\pdfoutput=1 % if your are submitting a pdflatex (i.e. if you have
\usepackage{jcappub} % for details on the use of the package, please
\usepackage[T1]{fontenc}

\usepackage{epsfig,latexsym,cancel,amssymb,amsmath}
\usepackage{hyperref}         % hypertext links %%ARXIV
\usepackage{graphicx}
\usepackage{times,psfrag,rotating}
\usepackage{aas_macros}

\usepackage{amsmath}
\usepackage{amssymb}
\usepackage{enumerate}
\usepackage{amsfonts}

\usepackage[caption=false]{subfig}
\usepackage{epstopdf}
\usepackage{color}
\usepackage{bm}
\usepackage[normalem]{ulem}
\usepackage{array}
\usepackage{multirow, multicol}

\renewcommand{\thetable}{$\arabic{table}$}

\newcolumntype{A}{ >{\centering\arraybackslash} m{3.5cm} }
\newcolumntype{B}{ >{\centering\arraybackslash} m{2.5cm} }
\newcolumntype{C}{ >{\centering\arraybackslash} m{2cm} }
\newcolumntype{D}{ >{\centering\arraybackslash} m{1.5cm} }
\newcolumntype{E}{ >{\centering\arraybackslash} m{2.5cm} }

\newcommand{\Neff}{$N_{\rm eff}$ }
\newcommand{\LCDM}{$\Lambda$CDM}

\def\beq{\begin{equation}}
\def\eeq{\end{equation}}
\def\bea{\begin{eqnarray}}
\def\eea{\end{eqnarray}}

\def\Eq#1{Eq.~(\ref{#1})}

\title{Alleviating the $H_0$ and $\sigma_8$  anomalies with a decaying dark matter model}

\affiliation[a]{International Centre for Theoretical Sciences, \\
Survey No. 151, Shivakote, Hesaraghatta Hobli, Bengaluru, Karnataka 560089, India}

\affiliation[b]{Department of Physics and Astronomy, Johns Hopkins University, \\
3400 N. Charles St., Baltimore, Maryland -- 21218, United States}

\affiliation[c]{Indian Institute of Astrophysics, \\
Sarjapura Road, 2nd Block Koramangala, Bengaluru, Karnataka -- 560034, India}

\author[a]{Kanhaiya L. Pandey,}
\emailAdd{kanhaiya.pandey@icts.res.in}

\author[b]{Tanvi Karwal}
\emailAdd{tkarwal@jhu.edu}

\author[c]{and Subinoy Das}
\emailAdd{subinoy@iiap.res.in}

\abstract{
The Hubble tension between the $\Lambda$CDM-model-dependent prediction of the current expansion rate $H_0$ using Planck data and direct, model-independent measurements in the local universe from the SH0ES collaboration disagree at $>3.5\sigma$. 
Moreover, there exists a milder $\sim 2\sigma$ tension between similar predictions for the amplitude $S_8$ of matter fluctuations and its measurement in the local universe. 
As explanations relying on unresolved systematics have not been found, theorists have been exploring explanations for these anomalies that modify the cosmological model, altering early-universe-based predictions for these parameters. 
However, new cosmological models that attempt to resolve one tension often worsen the other. 
In this paper, we investigate a decaying dark matter (DDM) model as a solution to both tensions simultaneously. 
Here, a fraction of dark matter density decays into dark radiation.
The decay rate $\Gamma$ is proportional to the Hubble rate $H$ through the constant $\alpha_{\rm dr}$, the only additional parameter of this model. 
Then, this model deviates most from $\Lambda$CDM in the early universe, with $\alpha_{\rm dr}$ being positively correlated with $H_0$ and negatively with $S_8$. 
Hence, increasing $\alpha_{\rm dr}$ (and allowing dark matter to decay in this way) can then diminish both tensions simultaneously. 
When only considering Planck CMB data and the local SH0ES prior on $H_0$,
$\sim 1$\% dark matter decays,
decreasing the $S_8$ tension to $0.3\sigma$ 
and increasing the best-fit $H_0$ by $1.6$ km/s/Mpc.
However, the addition of intermediate-redshift data (the JLA supernova dataset and baryon acoustic oscillation data) weakens the effectiveness of this model. 
Only $\sim 0.5$\% of the dark matter decays
bringing the $S_8$ tension back up to $\sim 1.5 \sigma$ and the increase in the best-fit $H_0$ down to $0.4$ km/s/Mpc. 
}

\begin{document}

%%%%%%%%%%%%%%%%%%%%%%%%%%%%%%%%%%%%%%%%%%%%%%%%%%%%%%%%%%%%%%%%%%%%%%%%%%%%%%%%%
%\pacs{95.35.+d,96.50.S-,98.80.Cq,95.85.Ry}
\maketitle

\newpage
%{\let\clearpage\relax\tableofcontents}
%%%%%%%%%%%%%%%%%%%%%%%%%%%%%%%%%%%%%%%%%%%%%%%%%%%%%%%%%%%%%%%%%%%%%%%%%%%%%%%%%

\section{Introduction}
\label{sec:intro}

The simple $\Lambda$CDM concordance model has been immensely successful in describing numerous cosmological observables at different epochs \cite{2016A&A...596A.107P,2016ApJ...826...56R,Ata:2017dya}. 
Nontheless, when fit to measurements of the early universe, the $\Lambda$CDM model finds results inconsistent with observations of the late universe \cite{Addison:2017fdm}. 
These include the persistent Hubble tension \cite{Freedman:2017yms} as well as the milder $S_8$ tension \cite{Battye:2014qga}. 

The current state-of-the-art experiment Planck which measures the cosmic microwave background (CMB) radiation, assumes a flat $\Lambda$CDM model to extract cosmological parameter values and finds the local expansion rate $H_0$ to be $67.37\pm 0.54$ km/s/Mpc \cite{2018arXiv180706209P}. 
On the other hand, the SH0ES collaboration finds the larger value $H_0 = 73.52\pm 1.62$ km/s/Mpc \citep[henceforth R18]{2018ApJ...861..126R} through model-independent measurements of the local universe, at $\gtrsim 3.5\sigma$ tension with the Planck value. 
This tension between the early and late universe exists even without Planck CMB data or the SH0ES distance ladder \cite{Addison:2017fdm}. 
Another direct measurement of $H_0 = 72.5^{+2.1}_{-2.3}$ km/s/Mpc \citep{Birrer:2018vtm} from the H0LiCOW collaboration based on lensing time delays is in moderate tension with Planck, 
while a constraint from Big Bang nucleosynthesis (BBN) combined with baryon acoustic oscillation (BAO) data of $H_0 = 66.98 \pm 1.18$ km/s/Mpc \cite{Addison:2017fdm} is inconsistent with SH0ES. 

There is also evidence of $\gtrsim 2\sigma$ tension between the constraints from Planck on the matter density $\Omega_m$ and the amplitude $\sigma_8$ of matter fluctuations in linear theory and those from local measurements \cite{Battye:2014qga,McCarthy:2017csu}. 
Planck derives $S_8 = \sigma_8 (\Omega_m/0.3)^{0.5} = 0.832\pm0.013$ whereas local measurements find the smaller values: 
$S_8^{\rm SZ}=\sigma_8(\Omega_m/0.27)^{0.3}=0.78\pm0.01$ from Sunyaev-Zeldovich cluster counts \citep{2014A&A...571A..20P}, 
$S_8=0.783^{+0.021}_{-0.025}$ from DES  \citep{2017arXiv170801530D}
and $S_8=0.745 \pm 0.039$ from KiDS-450 \cite{Hildebrandt:2016iqg} weak-lensing surveys. 
The CFHTLenS weak-lensing survey also finds support for disagreement with Planck CMB predictions \cite{Joudaki:2016mvz}. 

Although systematic causes for these discrepancies cannot entirely be ruled out, numerous potential systematics have been investigated and exonerated over the years while the tensions have persisted and worsened \cite{2018arXiv180706209P,McCarthy:2017csu,Efstathiou:2013via,Addison:2015wyg,Aghanim:2016sns,Shanks:2018rka,Riess:2018kzi,Soltis:2019ryf}. 
Hence, we must consider the alternative - that the model-dependent results from the early universe are inconsistent with the model-independent measurements of the late universe because the 
$\Lambda$CDM model of cosmology is incorrect. 

There have been numerous attempts at resolving these discrepancies via non-standard cosmological models 
\citep[and references therein]{2018arXiv181104083P,Karwal:2016vyq,2018arXiv180504716B,2018PhRvD..97l3504P,2017PhRvD..96b3523D,2016PhLB..761..242D,2015PhRvD..92l1302D,2007JCAP...05..008I,Enqvist:2015ara,DiValentino:2018gcu,2015PhRvD..92f1303B,2018JCAP...05..052N,2017JCAP...11..005A,2018PhRvD..98d3521K,2016PhRvD..94b3528C}, however, most such attempts at solving the Hubble tension worsen the $S_8$ tension and vice-versa. 
Solutions to the Hubble tension either reduce the size $r_s$ of the sound horizon with an early-universe modification \cite{2016JCAP...10..019B,Evslin:2017qdn,Aylor:2018drw}, 
or increase the angular diameter distance $D_A$ to the CMB with new physics in the post-recombination universe. 
Then, to keep the locations of the peaks in the CMB fixed, $H_0$ increases, diminishing the tension. 
On the other hand, a solution to the $S_8$ tension would require either late-universe physics that leads to a suppression of the linear matter power spectrum or a decrease in the CMB-predicted value of $\Omega_m$. 

In this paper, we tackle both requirements with a decaying dark matter (DDM) model which has a decay rate proportional to the Hubble rate. 
In this scenario, a fraction of dark matter density decays into dark radiation per Hubble time \citep{2012JCAP...10..017E}, 
with the effect being amplified close to the onset of matter domination. 
This leads to an increase in the expansion rate relative to $\Lambda$CDM around recombination, resulting in a decrease in $r_s$. 
Fits to the CMB then predict a higher $H_0$, alleviating the Hubble tension. 
These fits also predict smaller $\Omega_m$, leading to smaller $S_8$.
This model can hence simultaneously diminish both the Hubble and the $S_8$ tensions. 
Testing against various cosmological datasets, we find that this DDM model can provide a  better fit to some datasets and simultaneously alleviate the two aforementioned tensions, but not fully resolve them. 
We also find that at most, a fraction $f_{\rm dm} \lesssim 0.003$ of dark matter can decay into dark radiation in the light of recent Planck, supernova and BAO data, and an external prior on $H_0$ from R18. 
This paper is organised as follows.
A brief description of the model is given in Section~\ref{sec:ddm_model}, along with its effect on observables. 
In Section~\ref{sec:method}, we provide a detailed description of our analysis. 
Our results are presented in Section~\ref{sec:results} and discussed in Section~\ref{sec:discussion} where we also conclude.

\section{Decaying dark matter model}
\label{sec:ddm_model}

Motivation for exploring a DDM resolution to the Hubble tension comes from considering the effective radiation degrees of freedom \Neff \cite{2016ApJ...826...56R,2018PhRvD..97l3504P,2018arXiv180706209P,2013PhRvD..88b3511K,2012PhRvD..85f3513H,2017PhLB..768...12K,2016PhLB..762..462K}. 
Increasing the amount of radiation in the early universe such that $\Delta$\Neff $\sim 0.4-1$ has been shown to diminish the Hubble tension  \citep{2016ApJ...826...56R}. 
This extra radiation must be `dark' as the presence of an extra photon-like component is strongly constrained by both BBN and the CMB \citep{2016A&A...594A..13P}. 
A fourth, massive, sterile neutrino could provide such extra dark radiation however, the existence of such a particle is constrained by oscillation experiments \citep{2016PhRvL.117g1801A}. 

The scenario explored here follows the model proposed by Ref.~\cite{2012JCAP...10..017E}. 
It involves dark radiation interacting within the dark sector, in particular, 
a particle (beyond the framework of the standard model) decaying into an extra dark  radiation component. 
All the dark radiation in this scenario is a product of dark matter decay and forms a small fraction of the total dark matter density. 
The decay rate $\Gamma$ determines this fraction $f_{\rm dm}$ of dark matter energy density that decays into dark radiation. 
This fraction remains nearly constant over time after matter-radiation equality. 
If $f_{\rm dm}$ is large, it can alter the expansion rate as shown in Fig.~\ref{fig:mat_pow}, which we demonstrate leads to a higher predicted $H_0$.
Moreover, the decay naturally reduces the amount of dark matter in galaxies and clusters leading to smaller predicted values of $S_8$.
A brief description of the background dynamics of the model, contribution of dark radiation to $N_{\rm eff}$ and effect on observables is discussed in the following subsections. 

\subsection{Background dynamics}
\label{subsec:bg_dynamics}

A general coupling between dark matter and dark radiation can be described by the energy balance equations \cite{2008JCAP...07..020V}
\begin{eqnarray}
\dot\rho_{\rm \scriptscriptstyle dm} + 3 H\rho_{\rm \scriptscriptstyle dm} &=& -Q \\
\dot\rho_{\rm \scriptscriptstyle dr} + 3 H (1+w_{\rm \scriptscriptstyle dr}) \rho_{\rm \scriptscriptstyle dr} &=& Q
\end{eqnarray}
where $\rho_{\rm \scriptscriptstyle dm}$ and $\rho_{\rm \scriptscriptstyle dr}$ are the dark matter and dark radiation energy densities and $H = ̇\dot{a}/a$ is the Hubble rate, where $a$ is the scale factor and overdots denote derivatives with respect to conformal time. 
We also assume dark radiation has an equation of state $w_{\rm \scriptscriptstyle dr}=P_{\rm \scriptscriptstyle dr}/\rho_{\rm \scriptscriptstyle dr} = 1/3$. 
A positive rate of energy transfer $Q$ denotes the direction of energy transfer from dark matter to dark radiation. Non-zero values of $Q$ imply that dark matter no longer redshifts exactly as $a^{-3}$ nor dark radiation as $a^{-4}$. 
We adopt the covariant form of the energy-momentum transfer 4-vector introduced in \cite{2008JCAP...07..020V}
\begin{equation}
Q = \Gamma \rho_{\rm \scriptscriptstyle dm},
\end{equation}
where the exact form of the interaction rate $\Gamma$ depends on the details of the particle physics of the decay process. 

Many forms of $\Gamma$ have been studied in the literature \cite{Poulin:2016nat,Lesgourgues:2015wza,Buen-Abad:2017gxg,Chacko:2016kgg,Enqvist:2015ara}. 
Here, we explore the simple case where $\Gamma = \alpha H$, where $\alpha$ is a constant and $H$ is the Hubble rate. 
Although we do not model the particle physics resulting in $\Gamma \propto H$, we refer the reader to two fundamental particle physics motivations for such an interaction. 
As discussed in Section 5 of Ref.~\cite{2012JCAP...10..017E}, if dark matter is a coherently oscillating scalar field and decays into light fermions similar to the reheating mechanism, it can give rise to our DDM set up.
It may also arise in the model proposed by Ref.~\cite{Bringmann:2018jpr}, where a fraction of dark matter converts to dark radiation through late kinetic decoupling and Sommerfeld-enhanced dark matter annihilation. 
The mass ranges for dark matter particles in each of these models differ greatly. 
As our analysis here is phenomenological, our constraints are independent of the mass of the dark matter particle undergoing decay. 
Interpreting these results in the framework of a particular fundamental model can translate our constraints to particle mass and interaction cross-section. 

For $\Gamma = \alpha H$, the background evolution is readily solved  
\begin{eqnarray}
    \rho_{\rm \scriptscriptstyle dm} &=& \rho_{\rm \scriptscriptstyle dm,0} a^{-(3+\alpha_{\rm dr})} \\
    \rho_{\rm \scriptscriptstyle dr} &=& \rho_{\rm \scriptscriptstyle dr,0} a^{-3(1+w_{\rm \scriptscriptstyle dr})} + \frac{\alpha_{\rm dr}}{\alpha_{\rm dr}-3w_{\rm \scriptscriptstyle dr}} \rho_{\rm \scriptscriptstyle dm,0} a^{-3} (a^{-3w_{\rm \scriptscriptstyle dr}}-a^{-\alpha_{\rm dr}}) 
    ,
    \label{eq:rho_dr_bg}
\end{eqnarray}
where the subscript $0$ denotes values today. 
For $w_{\rm \scriptscriptstyle dr} = 1/3$, \Eq{eq:rho_dr_bg} can be further simplified to
\begin{equation}
    \rho_{\rm \scriptscriptstyle dr} = \beta a^{-4} + \frac{\alpha_{\rm dr}}{1-\alpha_{\rm dr}} \rho_{\rm \scriptscriptstyle dm,0} a^{-(3+\alpha_{\rm dr})} ,
    \label{eq:rho_dr_bg_simple}
\end{equation}
where $\beta$ is a constant. 
The first term in \Eq{eq:rho_dr_bg_simple} behaves like a standard radiation density while the second behaves like a fluid with an equation of state $\alpha_{\rm dr}/3$. 
For weak couplings between dark matter and dark radiation, $\alpha_{\rm dr} \ll 1$, which leads to $\beta \sim \rho_{\rm \scriptscriptstyle dr,0}$.
Assuming no dark radiation exists initially, we set $\beta = 0$ and only retain the second term in our analysis. 
With this assumption, we obtain the fraction $f_{\rm dm}$ of dark matter that decays into dark radiation 
\begin{equation}
    f_{\rm dm} = 
    \frac{\rho_{\rm \scriptscriptstyle dr}}{\rho_{\rm \scriptscriptstyle dm}} \rightarrow \frac{\alpha_{\rm dr}}{3w_{\rm \scriptscriptstyle dr}-\alpha_{\rm dr}} = \frac{\alpha_{\rm dr}}{1-\alpha_{\rm dr}}
    \simeq \alpha_{\rm dr}
    .
    \label{eq:frac_dr_dm}
\end{equation}
Therefore, $f_{\rm dm}$ is constant over time and the density of dark radiation $\rho_{\rm dr} \simeq \alpha_{\rm dr} \rho_{\rm dm}$. Our model is then parameterised by a single parameter $\alpha_{\rm dr}$. 

For a detailed description of the perturbations in our model, we refer the reader to Ref.~\cite{Wang:2012eka,2012JCAP...10..017E}.

\subsection{Calculation of $\Delta N_{\rm eff}$}
\label{subsec:N_eff}

In standard cosmology, the energy density $\rho_{\rm rad}$ of relativistic species in terms of the photon energy density $\rho_{\gamma}$ is
\begin{equation}
    \rho_{\rm r} = \left[ 1+ \frac{7}{8} N_{\rm eff} \left( \frac{4}{11}\right)^{4/3} \right] \rho_\gamma
    .
\end{equation}
This includes standard model (SM) neutrinos (for which $N_{\nu, \rm eff} = 3.046$) 
\citep{2005NuPhB.729..221M,2002PhLB..534....8M}, and characterizes  any  free-streaming  radiation  beyond  the SM  expectation. 
Then, any departure from the SM can be accounted for through $\Delta N_{\rm eff}$, where $N_{\rm eff} = N_{\nu,\rm eff}+\Delta N_{\rm eff}$. 
In our case, $\Delta N_{\rm eff,dr}$ can be expressed in terms of $\alpha_{\rm dr}$ \citep{2012JCAP...10..017E} as  
\begin{equation}
    \frac{7}{8} \Delta N_{\rm eff,dr} \left( \frac{4}{11} \right)^{4/3} \frac{\rho_{\scriptscriptstyle \gamma,0}}{a^4} = \frac{1}{1-\alpha_{\rm dr}} \rho_{\rm \scriptscriptstyle dm,0} a^{-(3+\alpha_{\rm dr})} - \rho_{\rm \scriptscriptstyle dm,0} a^{-3}
    ,
\end{equation}
making it a derived parameter in our analysis.

\subsection{Effects on observables}
\label{subsec:effect_observables}

\begin{figure}[h]
    \begin{center}
        \includegraphics[width=0.95\linewidth]{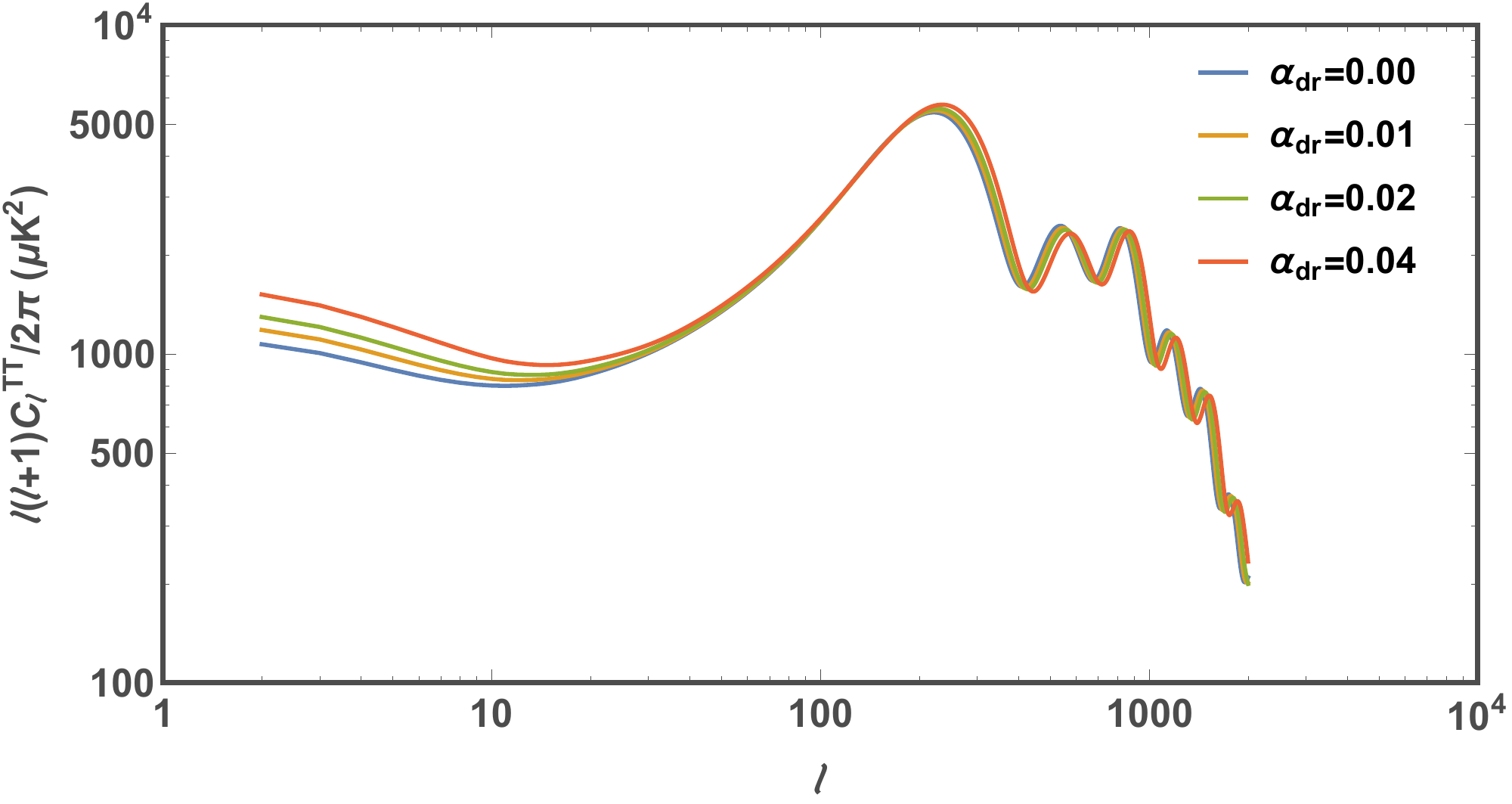}
        \includegraphics[width=0.45\columnwidth]{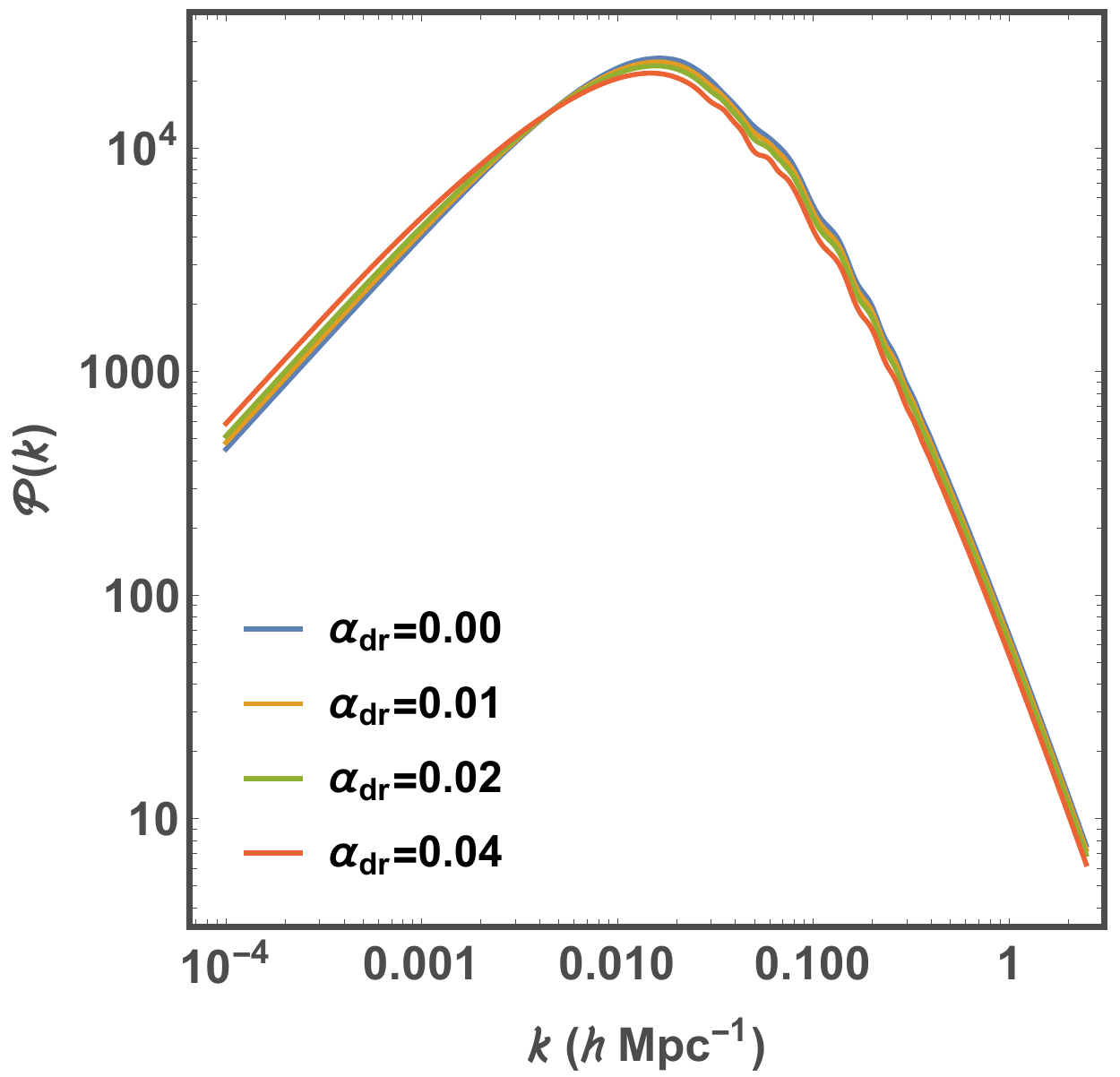}
        \includegraphics[width=0.45\columnwidth]{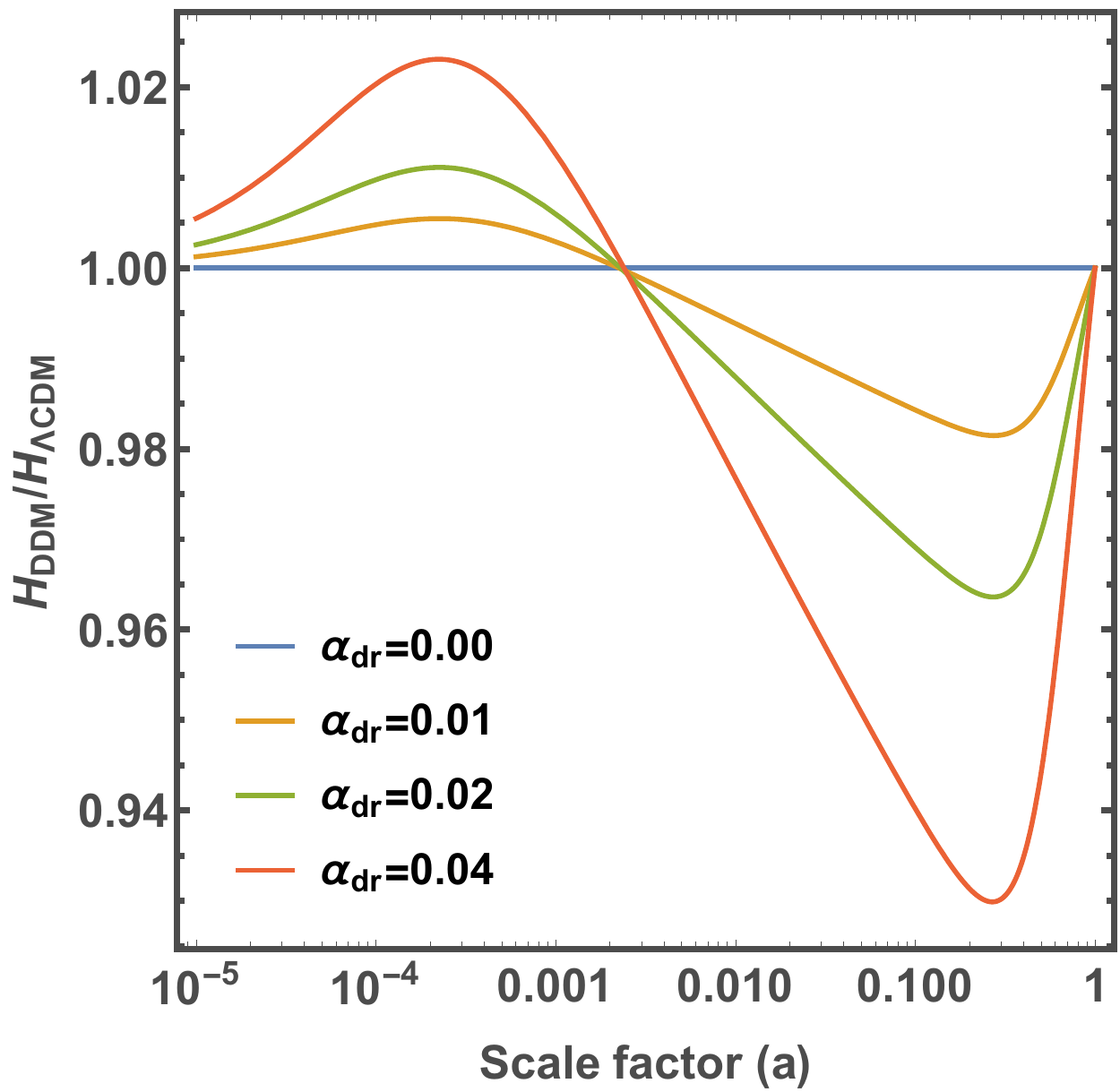}
        \caption {
        Shown here are the effects
        of DDM on various observables. 
        These plots were produced using a modified version of CAMB, 
        fixing all $\Lambda$CDM parameters except $\Omega_{\rm dm, 0}$ and deriving $\Omega_{\Lambda,0}$ by imposing flatness
        (see Section~\ref{subsec:effect_observables}).
        The blue line with $\alpha_{\rm dr} = 0$ represents a $\Lambda$CDM cosmology. 
        \textit{Top}: effect of non-zero $\alpha_{\rm dr}$ on the CMB TT power spectrum;
        \textit{left}: effect on the matter power spectrum;  
        \textit{right}: the DDM expansion rate relative to $\Lambda$CDM.
        }
        \label{fig:mat_pow}
    \end{center}
\end{figure}

To understand the effect of our model on observables, 
Fig.~\ref{fig:mat_pow} shows how decaying dark matter affects the CMB TT power spectrum, the matter power spectrum and the expansion rate. 
These plots were produced by fixing all \LCDM\ parameters except $\Omega_{\rm dm,0}$ to their best-fit \LCDM\ values, with $\Omega_{\Lambda,0} = 1 - \Omega_{\rm m,0}$ preserving flatness. 
The early universe fixes $\Omega_{\rm dm}$ at early times,
that is, if dark matter decays, $\Omega_{\rm dm,0}$ should decrease as its decay rate increases. 
This is also reflected in our posteriors, most evident in Fig.~\ref{fig:with_R18} from the negative correlation of $\Omega_{\rm dm,0}$ and $\alpha_{\rm dr}$.
Therefore, to better represent the cosmology in our posteriors, we fix matter-radiation equality at $a_{\rm eq}$ 
and hence $\Omega_{\rm dm}$ at early times and determine $\Omega_{\rm dm,0}$ as a function of $\alpha_{\rm dr}$. 
Then, at $a_{\rm eq}$, equating the matter and radiation contents of the Universe, 
\begin{eqnarray}
    \Omega_{\rm dm,0} a_{\rm eq}^{-(3+\alpha_{\rm dr})} +
    \Omega_{\rm b,0} a_{\rm eq}^{-3} = 
    \Omega_{\rm r,0} a_{\rm eq}^{-4} +
    \frac{\alpha_{\rm dr}}{1-\alpha_{\rm dr}} 
    \Omega_{\rm dm,0} a_{\rm eq}^{-(3+\alpha_{\rm dr})} 
    ,
\end{eqnarray}
where we have substituted $\rho_{\rm dr}$ using Eq.~\eqref{eq:frac_dr_dm}. 
Then for the dark sector today, we have 
\begin{eqnarray}
    \Omega_{\rm dm,0} & = &
    \frac{a_{\rm eq}^{-1}\Omega_{\rm r} - \Omega_{\rm b}}
    {a_{\rm eq}^{-\alpha_{\rm dr}}} 
    \left( \frac{1-\alpha_{\rm dr}}{1-2\alpha_{\rm dr}} \right) 
    , \mbox{\rm\ and}
    \\
    \Omega_{\rm \Lambda,0} & = &
    1 - \Omega_{\rm dm,0} - \Omega_{\rm b}.
\end{eqnarray}

The main effect of the DDM model is an alteration of the expansion history of the Universe, as seen from the bottom-right panel of Fig.~\ref{fig:mat_pow}. 
At very early times, the Universe is radiation-dominated and the DDM universe resembles the \LCDM\ universe as dark radiation is always subdominant.  
As we fix $\Omega_{\rm dm}$ at $a_{\rm eq}$, the pre-equality DDM universe has more dark matter than \LCDM, increasing its expansion rate with $H_{\rm DDM}/H_{\rm \Lambda CDM}$ peaking at $a_{\rm eq}$. 
After $a_{\rm eq}$, the DDM universe has less dark matter than \LCDM\ 
as dark matter decays into dark radiation which quickly redshifts away
and $H_{\rm DDM}$ decreases. 
The two expansion rates are equal when $\Omega_{\rm dm} + \Omega_{\rm dr} = \Omega_{\rm cdm}$ which occurs at
\begin{eqnarray}
      a_{\rm cross} = \frac{1}{(1- 2 \alpha_{\rm dr})^{(1/\alpha_{\rm dr})}} a_{\rm eq}
      ,
      \label{eq:a_equal_H}
\end{eqnarray}
ignoring minor variations due to $\Omega_{\Lambda, 0}$. 
The DDM expansion rate continues to decrease past this point until $\Lambda$ becomes important and begins to dominate. 
The Universe again resembles a \LCDM\ universe and $H_{\rm DDM} = H_{\rm \Lambda CDM}$ at $a=1$ as set by our choice of $\Omega_{\Lambda,0}$. 
Hence, the early, pre-recombination universe expands faster than \LCDM\ and the post-recombination universe slower. 
This effect is more pronounced as $\alpha_{\rm dr}$ increases. 

The effect of DDM on the CMB is shown in the top panel of Fig.~\ref{fig:mat_pow}.
The increase in the pre-recombination expansion rate and decrease in the post-recombination expansion rate lead to a smaller $r_s$ and larger angular distance $D_A$ to the CMB respectively. 
The inferred angular size of the sound horizon $\theta_* = r_s / D_A$ is then smaller and the CMB peaks are shifted to smaller scales or larger multipoles $\ell$ as $\alpha_{\rm dr}$ increases.
These can be shifted back into agreement with data by increasing the Hubble constant, relieving the tension. 
Other changes to the CMB include the following suppressions and enhancements of power. 
An increase in the amount of dark matter in the pre-equality universe suppresses power in CMB peaks 
which entered the horizon during radiation domination, 
as the enhancement due to acoustic driving is reduced.
On the other hand, as there is less dark matter post-equality, 
the first CMB peak receives a boost in power, 
having entered the horizon during matter-domination.
Finally, $\Omega_{\Lambda,0}$ is larger for larger $\alpha_{\rm dr}$ and the late universe is $\Lambda$-dominated sooner. 
The late ISW effect hence enhances power in low multipoles. 
Small shifts in the other \LCDM\ parameters can absorb these changes in the CMB. 

Finally, the matter power spectrum, shown in the bottom-left panel of Fig.~\ref{fig:mat_pow} is effectively inflected about $k_{\rm cross} \simeq a_{\rm cross}H$ where $a_{\rm cross}$ is given by Eq.~\eqref{eq:a_equal_H}. 
At scales larger than $k_{\rm cross}$, there is more power in a DDM universe for larger $\alpha_{\rm dr}$, and less at scales smaller than $k_{\rm cross}$. 
Relating to the bottom-right plot in Fig.~\ref{fig:mat_pow} of $H_{\rm DDM}/H_{\rm \Lambda CDM}$, 
we find that the change to the expansion rate dictates the power in the matter power spectrum. 
At early times, when the DDM universe is expanding faster than the \LCDM\ universe, clustering and therefore power in the matter power spectrum is suppressed. 
The opposite is true after $a_{\rm cross}$, or below $k_{\rm cross}$ - the DDM universe expands slower than \LCDM\ and power in the matter power spectrum is boosted. 
These trends are enhanced for larger values of $\alpha_{\rm dr}$, as expected. 
The BAO peaks at $k > k_{\rm cross}$ are also slightly shifted to smaller scales, same as the trend in $r_s$ in the CMB, and sharper as the pre-equality universe has more dark matter. 
Altogether, as power is suppressed at scales $\sim 8$Mpc, the DDM model helps alleviate the $S_8$ tension.

\section{Methodology}
\label{sec:method}

To investigate this DDM model, we use a modified version of the publicly available Boltzmann code \verb|CAMB| \citep{2000ApJ...538..473L}. 
The modified version is based on the dynamics described in  \citep{2012JCAP...10..017E}; 
We vary the 6 standard $\Lambda$CDM parameters: 
the baryon density $\omega_{\rm b}$ today,
the dark matter density $\omega_{\rm dm}$ today, 
the angular size $\theta_{\rm MC}$ of the sound horizon at recombination, 
the optical depth $\tau$ to reionisation, 
the scalar spectral index $n_{\rm s}$,
and the amplitude $A_{\rm s}$ of the primordial power spectrum. 
To this, we add the DDM parameter $\alpha_{\rm dr}$. 
We then use the publicly available Markov chain Monte Carlo code \verb|CosmoMC| \citep{2013PhRvD..87j3529L,2002PhRvD..66j3511L} to explore our 7-dimensional parameter space with the following assumptions. 
We assume a flat universe with $\Omega_k = 0$, 
a constant dark energy equation of state, $w_{\rm de}=-1$ and fix the running of the scalar spectral index $dn_{\rm s}/d{\rm ln}k = 0$. 
We adopt standard values for the sum of neutrino masses $\Sigma m_\nu = 0.06$ eV and the SM  $N_{\nu,\rm eff} = 3.046$. 
The entire DDM model is described by the sole parameter $\alpha_{\rm dr}$. 
The dark radiation energy density $\Omega_{\rm dr}$ and $\Delta N_{\rm eff,dr}$ are derived parameters which can be expressed in terms of $\alpha_{\rm dr}$. 
Table~\ref{tab:priors} shows the priors for the 7 varied parameters. 

We fit to various early and late-universe data sets in certain combinations. Our data include: 
\begin{itemize}
\item {\bf Planck :}  The CMB temperature and polarization angular power spectra  (high-$\ell$ TT + low-$\ell$ TEB) released by Planck 2015 \cite{2016A&A...594A..13P,2016A&A...594A..11P}
\item {\bf JLA :}  Luminosity distance of supernovae Type Ia coming from `joint light-curve analysis' using SNLS (Supernova Legacy Survey) and SDSS (Sloan Digital Sky Survey) catalogs \citep{2014A&A...568A..22B}
\item {\bf BAO :}  The `Baryonic Acoustic Oscillation' data from DR12-BAO \citep{2017MNRAS.470.2617A}, SDSS-6DF \citep{2011MNRAS.416.3017B} and SDSS-MGS \citep{2015MNRAS.449..835R}
\item {\bf R18 :}  An external gaussian prior on $H_0 = 73.52\pm1.62$ km/s/Mpc \citep{2018ApJ...861..126R}.
\end{itemize}
We fit to the combinations Planck, Planck+R18
and Planck+JLA+BAO+R18.
We adhere to the Gelman-Rubin convergence criteria of $R-1<0.01$ and discard the first 30\% of our chains as burn-in.

\begin{table}[!h]
\begin{center}
\renewcommand{\arraystretch}{1.2}
\scalebox{1.0}{
\begin{tabular} {|l|c|c|}
\hline
\hline
	Parameter & $\Lambda$CDM & DDM \\
\hline
$\Omega_b h^2                 $ & [  0.005 , 0.1  ] & [  0.005 , 0.1  ] \\
$\Omega_{dm} h^2                 $ & [  0.001 , 0.99 ] & [  0.001 , 0.99 ] \\
$100\theta_{MC}               $ & [  0.5   , 10   ] & [  0.5   , 10   ] \\
$\tau                         $ & [  0.01  , 0.8  ] & [  0.01  , 0.8  ] \\
$n_s                          $ & [  0.8   , 1.2  ] & [  0.8   , 1.2  ] \\
${\rm{ln}}(10^{10} A_s)       $ & [  2.0   , 4.0  ] & [  2.0   , 4.0  ] \\                                                               
$\alpha_{\rm dr}                  $ & --                & [  0.00  , 0.05 ] \\
\hline                                                   
\hline
\end{tabular}
}
\caption{Priors on the cosmological parameters we vary in our MCMC analyses}
\label{tab:priors}
\end{center}
\end{table}

\section{Results}
\label{sec:results}

\begin{figure}
\begin{center}
\includegraphics[width=1.0\columnwidth]{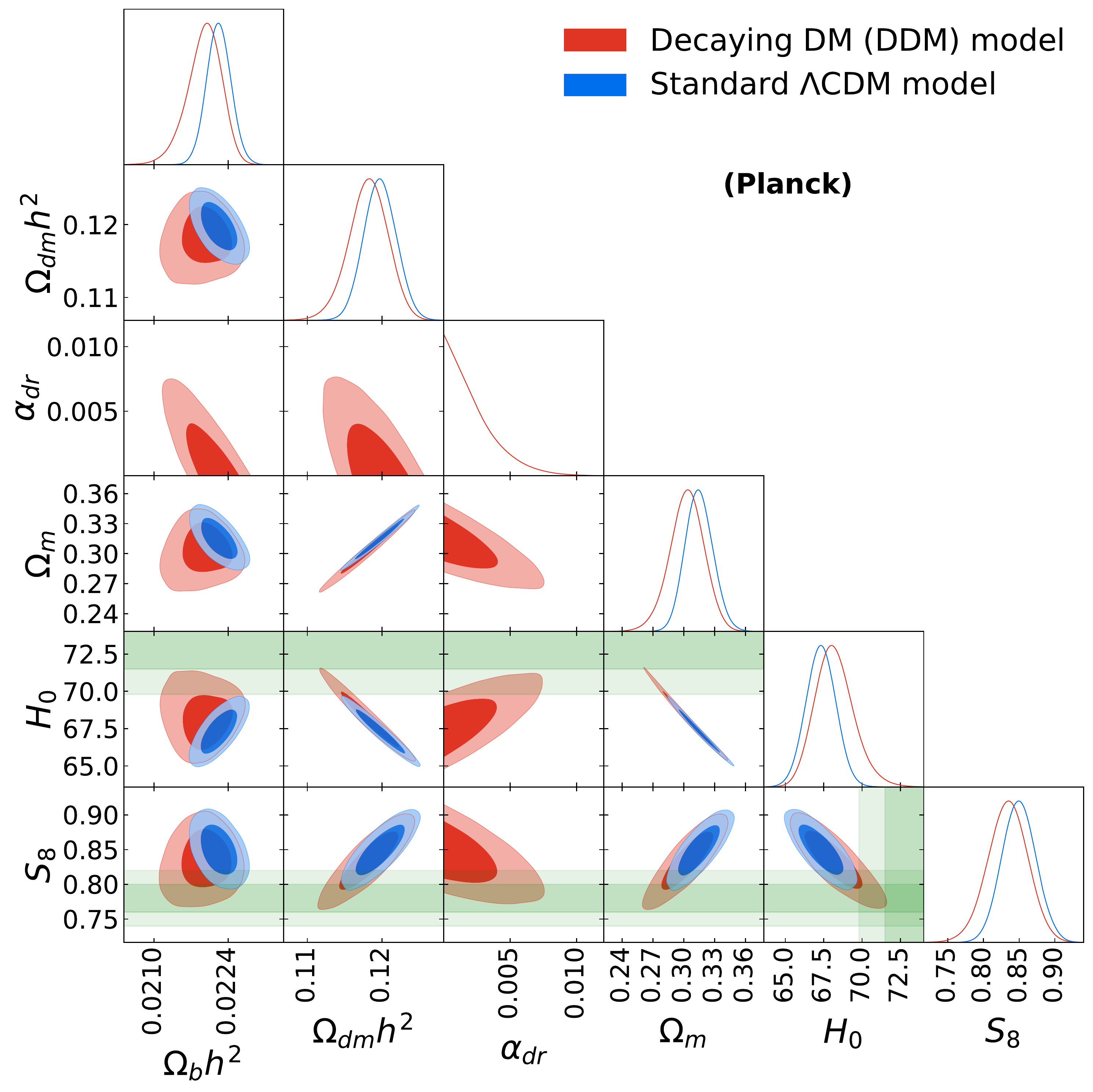}
\caption {Comparison between the standard $\Lambda$CDM and the DDM models: Constraints on various cosmological parameters along with their covariances when tested against the Planck data. The green bands represent the constraints on $H_0$ and $S_8$ coming from \citep[R18]{2018ApJ...861..126R} and \citep[DES-YI, 2017]{2017arXiv170801530D}. 
The positive correlation between $H_0$ and $\alpha_{\rm dr}$ and the negative correlation $S_8$ and $\alpha_{\rm dr}$ can be seen here. 
}
\label{fig:Planck_only}
\end{center}
\end{figure}

\begin{figure}
\begin{center}
\includegraphics[width=1.0\columnwidth]{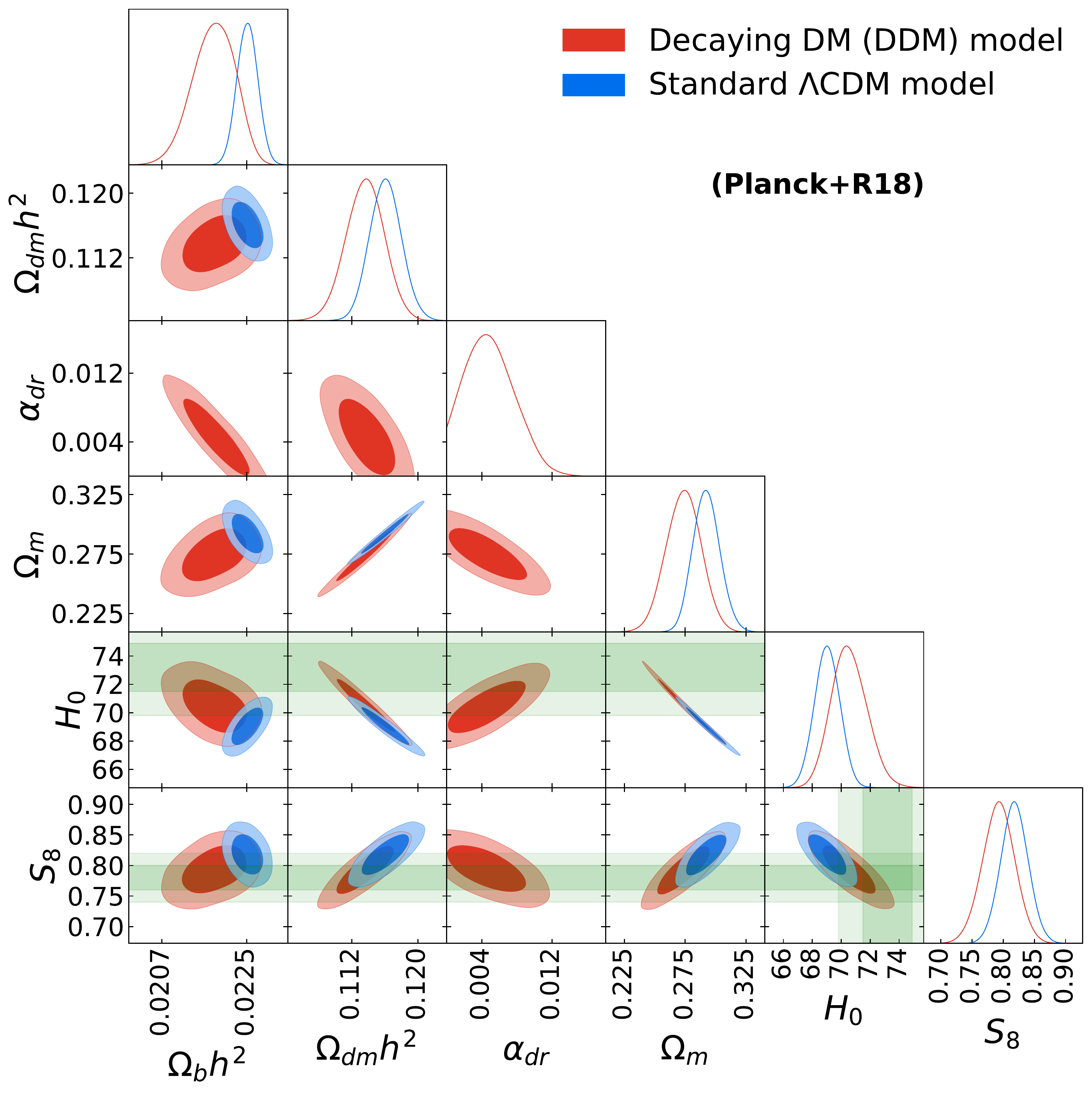}
\caption {
Comparison between the standard $\Lambda$CDM and the DDM models: Constraints on various cosmological parameters along with their covariances when tested against the Planck+R18. The green bands represent the constraints on $H_0$ and $S_8$ coming from \citep[R18]{2018ApJ...861..126R} and \citep[DES-YI, 2017]{2017arXiv170801530D}.
The correlations of $\alpha_{\rm dr}$ with $H_0$ and $S_8$ are more clearly visible here from the tilts of their contours. 
This data set combination also has the largest shift in $\Omega_m$, which helps relieve both tensions. 
}
\label{fig:with_R18}
\end{center}
\end{figure}

\begin{figure}
\begin{center}
\includegraphics[width=1.0\columnwidth]{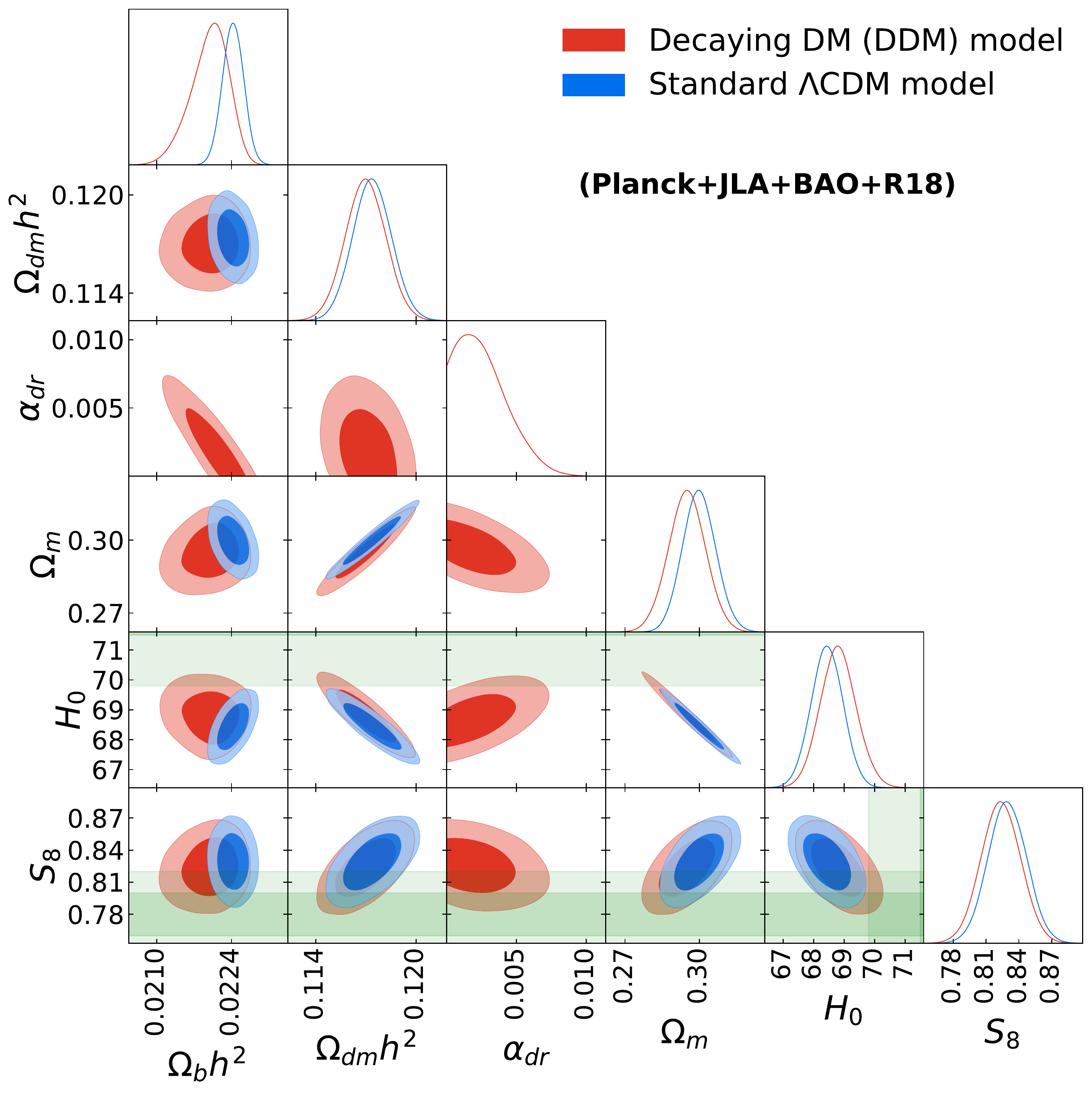}
\caption {Comparison between the standard $\Lambda$CDM and the DDM models: Constraints on various cosmological parameters along with their covariances when tested against the Planck+JLA+BAO+R18. The green bands represent the constraints on $H_0$ and $S_8$ coming from \citep[R18]{2018ApJ...861..126R} and \citep[DES-YI, 2017]{2017arXiv170801530D}.}
\label{fig:with_JLA+BAO+R18}
\end{center}
\end{figure}

Figures~\ref{fig:Planck_only}-\ref{fig:with_JLA+BAO+R18} compare our posteriors for the $\Lambda$CDM (blue) and DDM (red) models for various data sets. 
Along with posteriors for $\Omega_b h^2$, $\Omega_{dm} h^2$ and $\alpha_{\rm dr}$, we also show posteriors for the derived parameters 
$H_0$ and $S_8$. 
The green bands represent local measurements of $H_0$ and $S_8$.
From these figures, the correlation of $H_0$ with $\alpha_{\rm dr}$ and the anticorrelation of $S_8$ with $\alpha_{\rm dr}$ are apparent. 
An increase in $\alpha_{\rm dr}$ results in a greater $H_0$ and a smaller $S_8$. 
This is the exact effect required to solve the $H_0$ and $S_8$ tensions simultaneously.  
These correlations are most evident in the posteriors of Planck+R18 in Fig.~\ref{fig:with_R18}.
The inclusion of JLA and BAO data diminishes these correlations as seen in Fig.~\ref{fig:with_JLA+BAO+R18}. 

As seen from Fig.~\ref{fig:Planck_only}, Planck data places an upper bound on $\alpha_{\rm dr}$ ($\leq0.003$). 
However, the addition of an external prior on $H_0$ from R18 leads to a small preference for non-zero $\alpha_{\rm dr}$ ($\approx 0.005\pm0.003$) at the $\sim 1.5\sigma$ level. 
With Planck+R18, the Hubble tension is reduced to  $\sim 1.5 \sigma$ and the $S_8$ tension to $\sim 0.3 \sigma$. 
The addition of JLA and BAO data weakens these resolutions, as seen form Table~\ref{tab:with_Planck+R18_Everything}. 
For Planck+JLA+BAO+R18, the $H_0$ and $S_8$ tensions remain at $\sim 2.5 \sigma$ and $\sim 1.5 \sigma$ levels respectively.
For all dataset combinations explored, we remain consistent with $\Lambda$CDM within $1\sigma$ for all $\Lambda$CDM parameters. 

Table~\ref{tab:chi_square} shows the best-fit $\chi^2$ values for the $\Lambda$CDM and DDM models. 
The DDM model leads to a slight improvement in fit, largely due to fitting the R18 measurement better than $\Lambda$CDM.

\begin{figure}
    \begin{center}
        \includegraphics[width=0.49\columnwidth]{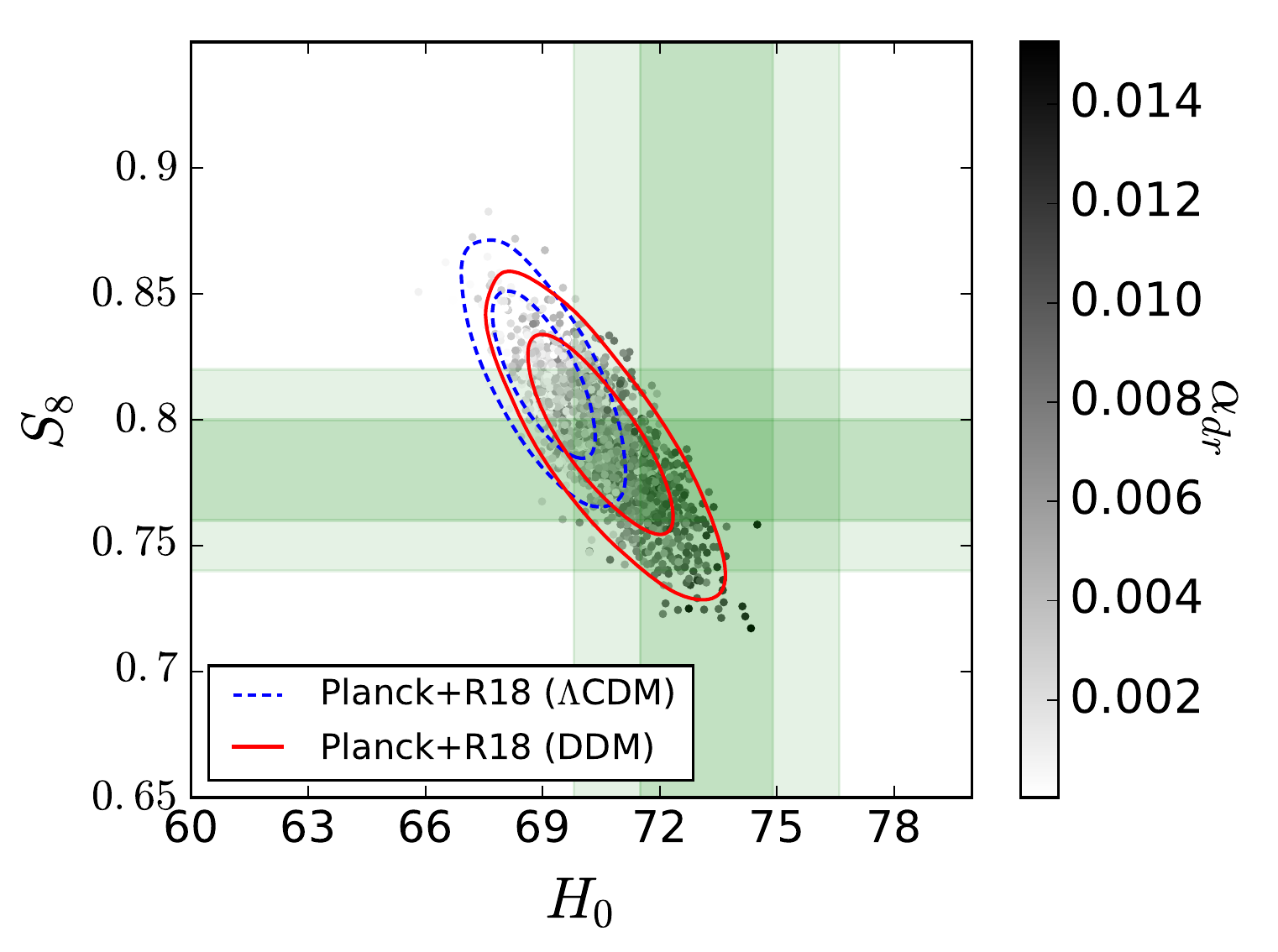}
        \includegraphics[width=0.49\columnwidth]{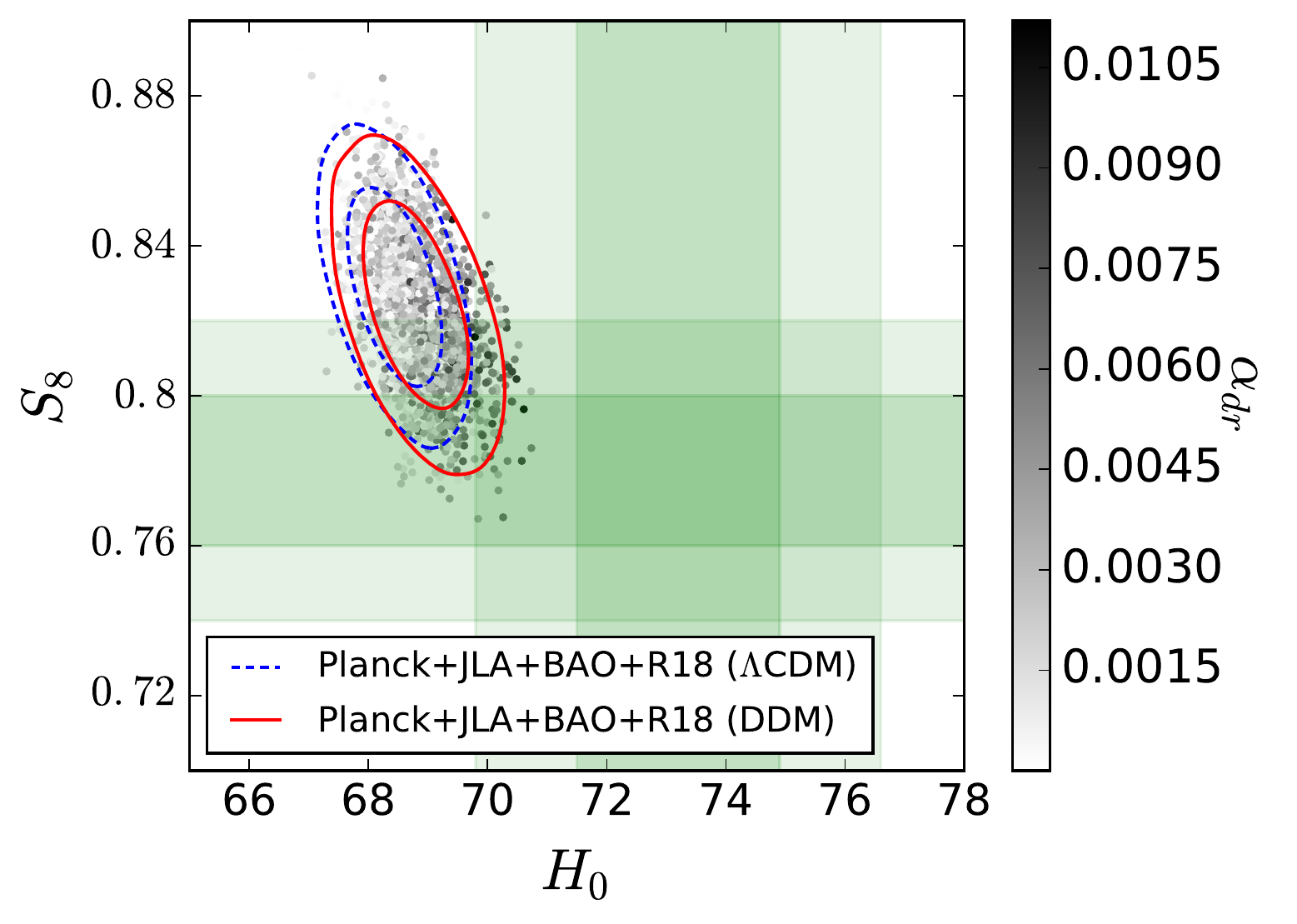}
        \caption {
            The figures show the $\Lambda$CDM (blue) and DDM (red) constraints on the $H_0-S_8$ plane for two dataset combinations, Planck+R18 and Planck+R18+JLA+BAO.
            The green bands represent the 1 and 2 $\sigma$ constraints on $H_0$ and $S_8$ coming from \citep[R18]{2018ApJ...861..126R} and \citep[DES-YI, 2017]{2017arXiv170801530D}. 
            The scattered points are for the DDM model representing values of $\alpha_{\rm dr}$.
        }
        \label{fig:3d_planck}
    \end{center}
\end{figure}

\clearpage
In Fig.~\ref{fig:3d_planck}, we show how the decay parameter $\alpha_{\rm dr}$ improves the $H_0$ and $S_8$ tensions. 
In the DDM scenario (red contours), $\alpha_{\rm dr}$ increases towards the bottom right. 
Considering just Planck+R18, the external prior on $H_0$ pushes the decay rate of dark matter to be $\sim 1\%$ of the Hubble rate. 
These large values of $\alpha_{\rm dr}$ alter expansion history enough for the model to predict a larger $H_0$. 
As the early universe fixes the density of dark matter, the DDM model also leads to smaller $\Omega_{dm}$ and therefore $\Omega_m$, lowering the predicted $S_8$. 
Without data at intermediate redshifts, such large changes in cosmology are permitted. 
As seen from the left panel of Fig.~\ref{fig:3d_planck}, within the scope allowed by Planck+R18, the DDM contours intersect the $1\sigma$ local-measurement square (green). 
For these datasets, while the $1\sigma$ $\Lambda$CDM contour (blue) intersects the $1\sigma$ local measurement of $S_8$, the $2\sigma$ $\Lambda$CDM contour is beyond the $1\sigma$ local $H_0$ measurement. 
Therefore, the DDM model diminishes the $H_0$ and $S_8$ tensions. 

Including data at intermediate redshifts, primarily supernova data from JLA, the tensions remain unresolved. 
As seen from the right panel of Fig.~\ref{fig:3d_planck}, while DDM nearly resolves the $S_8$ tension, the Hubble tension still exists. 
The combined datasets Planck+R18+JLA+BAO do not permit large deviations from $\Lambda$CDM cosmology, 
which is in agreement with  other studies where it has been found that supernovae and BAO measurements generically prefer standard $\Lambda$CDM \citep{2017PhRvD..96b3523D,2018PhRvD..97h3508C}.
Moreover, smaller values of $\alpha_{\rm dr}$ are permitted, with the decay rate of dark matter constrained to be $\sim 0.5\%$ of the Hubble rate.

\begin{table}
\renewcommand{\arraystretch}{1.2}
\begin{tabular} {| l | c c | c c|}
\hline
 &  
 \multicolumn{2}{c|}{Planck+R18} &  \multicolumn{2}{c|}{Planck+JLA+BAO+R18}\\
\hline
 Parameter & $\Lambda$CDM &  DDM & $\Lambda$CDM &  DDM\\
\hline
$\Omega_b h^2          $ & $0.02251\pm 0.00022      $ & $0.02180^{+0.00049}_{-0.00041} $ & $0.02243\pm 0.00020 $ & $0.02199^{+0.00041}_{-0.00029}$ \\
$\Omega_{dm} h^2       $ & $0.1161\pm 0.0019        $ & $0.1136\pm 0.0023              $ & $0.1174\pm 0.0012   $ & $0.1170\pm 0.0012             $ \\
$100\theta_{MC}        $ & $1.04138\pm 0.00045      $ & $1.04118\pm 0.00045            $ & $1.04123\pm 0.00041 $ & $1.04098\pm 0.00044           $ \\
$\tau                  $ & $0.094\pm 0.019          $ & $0.089\pm 0.020                $ & $0.089\pm 0.018     $ & $0.082\pm 0.019               $ \\
${\rm ln}(10^{10} A_s) $ & $3.113\pm 0.037          $ & $3.107\pm 0.038                $ & $3.106\pm 0.035     $ & $3.096\pm 0.037               $ \\
$n_s                   $ & $0.9748\pm 0.0058        $ & $0.9763\pm 0.0058              $ & $0.9715\pm 0.0043   $ & $0.9699\pm 0.0044             $ \\
$\alpha_{\rm dr}       $ & $--                    	$ & $0.0050^{+0.0023}_{-0.0034}    $ & $--                 $ & $< 0.00332                    $ \\
\hline                                                                                                                                                 
$\Omega_m              $ & $0.293\pm 0.011          $ & $0.274\pm 0.014                $ & $0.3000\pm 0.0067   $ & $0.2950\pm 0.0074             $ \\
$\Omega_\Lambda        $ & $0.707^{+0.012}_{-0.010} $ & $0.725\pm 0.014                $ & $0.7000\pm 0.0067   $ & $0.7044\pm 0.0073             $ \\
$\sigma_8              $ & $0.829\pm 0.015          $ & $0.830\pm 0.015                $ & $0.830\pm 0.015     $ & $0.831\pm 0.015               $ \\
$S_8                   $ & $0.818\pm 0.022          $ & $0.793\pm 0.026                $ & $0.829\pm 0.018     $ & $0.823\pm 0.018               $ \\
$H_0                   $ & $69.03\pm 0.87           $ & $70.6^{+1.1}_{-1.3}            $ & $68.44\pm 0.52      $ & $68.81\pm 0.58                $ \\
\hline
\end{tabular}
\caption {Comparison between the standard $\Lambda$CDM and the DDM models showing $1\sigma$ constraints on parameters fitting to Planck+R18 and Planck+JLA+BAO+R18}
\label{tab:with_Planck+R18_Everything}
\end{table}

\begin{table}
\begin{center}
\renewcommand{\arraystretch}{1.2}
\begin{tabular} {| l | c c | c c|}
\hline
 &  
 \multicolumn{2}{c|}{Plank+R18} &  \multicolumn{2}{c|}{Plank+JLA+BAO+R18}\\
\hline
 Dataset & $\Lambda$CDM &  DDM & $\Lambda$CDM & DDM \\
\hline
$ \chi^2_{\rm high\ell TT}       $ & $768.352   $ & $771.684    $ & $767.395   $ & $767.154   $\\
$ \chi^2_{\rm lowTEB}            $ & $10498.3   $ & $10496.5    $ & $10497.3   $ & $10497.9   $\\
$ \chi^2_{\rm JLA}               $ & $--        $ & $--         $ & $695.377   $ & $695.299   $\\
$ \chi^2_{\rm 6DF}               $ & $--        $ & $--         $ & $0.0402244 $ & $0.0793526 $\\
$ \chi^2_{\rm MGS}               $ & $--        $ & $--         $ & $2.34994   $ & $2.67358   $\\
$ \chi^2_{\rm DR12BAO}           $ & $--        $ & $--         $ & $3.57457   $ & $3.96993   $\\
$ \chi^2_{\rm nuisance}             $ & $1.50061   $ & $3.20412    $ & $3.11594   $ & $2.42671   $\\
$ \chi^2_{\rm R18}               $ & $7.59589   $ & $2.46971    $ & $9.11007   $ & $7.98282   $\\
\hline
$ \sum\chi^2_{\rm i}           $ & $11275.7   $ & $11273.8    $ & $11978.3   $ & $11977.5   $\\
$ \Delta(\sum\chi^2_{\rm i})     $ & $0         $ & $-1.9       $ & $0         $ & $-0.8      $\\
\hline
\end{tabular}
\caption {Comparison between the standard $\Lambda$CDM and the DDM models: $\chi^2$ values for various datasets from a combined fit to Planck+R18 and Planck+JLA+BAO+R18 are given, with the $\chi^2_{\rm nuisance}$ for expectations for the nuisance and foreground parameters.}
\label{tab:chi_square}
\end{center}
\end{table}

\clearpage

\section{Discussion and conclusions}
\label{sec:discussion}

While the $\Lambda$CDM model of cosmology fits numerous datasets well, its predictions based on the early and late universe disagree \cite{Addison:2017fdm,Aylor:2018drw}. 
The current expansion rate $H_0$ is underpredicted by $\Lambda$CDM when fit to the early universe \cite{2018ApJ...861..126R,2018arXiv180706209P}. 
Despite this, measurements of the late universe are in agreement with a $\Lambda$CDM expansion history, but with different parameter values \cite{2016ApJ...826...56R}. 
This Hubble tension has persisted and worsened over the years and no systematic cause has yet been found \cite{Freedman:2017yms}.
Moreover, $\Lambda$CDM overpredicts the amplitude of matter fluctuations $S_8$ relative to direct measurements in the late universe \cite{Battye:2014qga,McCarthy:2017csu}. 
Although this is a milder tension, combined, these tensions might hint new physics beyond the standard model of cosmology. 

Theories that address each tension often worsen the other. 
In this paper, we explored a decaying dark matter model than can simultaneously improve both tensions.
We considered dark matter that decays into dark radiation, parameterised by a single new parameter \cite{2008JCAP...07..020V,2012JCAP...10..017E}.
The DDM model increases the expansion rate relative to $\Lambda$CDM, with the largest effect being close to recombination. 
This leads to a reduced sound horizon, to compensate for which $H_0$ increases, alleviating the Hubble tension. 
The DDM model also reduces the dark matter density in the late universe, suppressing structure formation and lowering the predicted value of $S_8$. Hence, it offers solutions to both tensions simultaneously. 

Considering just data from the early and current universes, that is the Planck+R18 combination, we find that the Hubble tension is reduced below the $1.5\sigma$ level and the $S_8$ tension below $0.5\sigma$. 
DDM not only significantly diminishes both tensions, but also provides a slightly better fit to these datasets with $\Delta \chi^2_{\rm tot} = -1.9$, as seen from Table~\ref{tab:chi_square}. 
However, including measurements of the Universe at intermediate redshifts with Planck+R18+JLA+BAO, we find that the DDM model is strongly constrained and the $H_0$ and $S_8$ tensions persist at the $\sim 2.5 \sigma$ and $\sim 1.5 \sigma$ levels respectively. 
The DDM model alters expansion history relative to $\Lambda$CDM all through matter domination, as shown in Fig.~\ref{fig:mat_pow}. 
As found by numerous models that aim to resolve the Hubble tension through modifications of the late universe, late-universe datasets such as JLA and BAO strongly constrain expansion history and keep such models from fully resolving the Hubble tension \cite{2018arXiv180706209P,2018PhRvD..97l3504P,2017PhRvD..96b3523D,2016PhLB..761..242D}. 
In this case, the ``new physics'' we add is present not only in the early universe where it has maximal effect, but throughout cosmic history. 
Its presence in the late universe would spoil the fits to JLA and BAO, keeping it from diminishing the $H_0$ and $S_8$ tensions further. 
This can also be seen from the tilt of the $H_0$ and $S_8$ vs $\alpha_{\rm dr}$ contours in Figs.~\ref{fig:with_R18} and \ref{fig:with_JLA+BAO+R18}. 
Without JLA and BAO data, $\alpha_{\rm dr}$ has a stronger correlation with $H_0$ and anticorrelation with $S_8$ in Fig.~\ref{fig:with_R18}. 
This relaxes when intermediate-redshift datasets are added as in Fig.~\ref{fig:with_JLA+BAO+R18}, implying that the addition of JLA and BAO data weakens the effectiveness of DDM at resolving both tensions. 

Numerous models of dark matter interacting within the dark sector have been explored \cite{Serra:2009uu,Cyr-Racine:2013fsa, Das:2017nub,Wilkinson:2014ksa}. 
In these models, the interaction is effective only up to a certain scale and negligible at larger scales. 
This produces a cut-off-like feature in the matter power spectrum at small scales, keeping the power in scales $\sim 8$Mpc the same as in $\Lambda$CDM. 
For decaying dark matter with a constant time-independent decay rate \cite{Enqvist:2015ara,Poulin:2016nat}, the constraints are driven by the change to the late integrated Sachs-Wolfe (ISW) effect on the large-scale CMB data. 
To be consistent with it, the dark matter must decay very slowly which only allows a slight improvement in the $S_8$ tension. 
The DDM model considered here circumvents this by having a smaller decay rate in the early universe around decoupling which then increases with time. 
Models which introduce a time-dependent dark-matter drag force due to dark radiation which also shut-off at late times \cite{Buen-Abad:2017gxg} have similar effects.
 % and can address the $S_8$ tension effectively. 

The $S_8$ and Hubble tensions are intriguing results in cosmology. 
They require careful investigation whether from a systematic or a new-physics perspective. 
Future data may shed further light on whether these anomalies are hints of physics beyond the standard model of cosmology after all.

\acknowledgments
We thank  Marc Kamionkowski and Vivian Poulin for their comments on the initial results of this work. 
SD and TK acknowledge the IUSSTF-JC-009-2016 award from the Indo-US Science \& Technology Forum which supported the project and facilitated the authors visiting each other. 
TK also acknowledges support from the 2018 Johns Hopkins Discovery Award.

%\newpage
%\appendix
%\section*{Appendix}

%%%%%%%%%%%%%%%%%%%%%%%%%%%%%%%%%%%%%%%%%%%%%%%%%

\bibliographystyle{utphys}

\bibliography{bibliography}{}
%%%%%%%%%%%%%%%%%%%%%%%%%%%%%%%%%%%%%%%%%%%%%%%%%

\end{document}